\documentclass[twocolumn,showpacs,preprintnumbers,amsmath,amssymb]{revtex4-1}

\usepackage{afterpage}
\usepackage[dvipdfmx]{graphicx}
\usepackage{color,bm}
\usepackage{tabularx}

\begin{document}

\title{
Gravitational lens on a static
optical constant-curvature background: Its application to Weyl gravity model}

\author{Keita Takizawa}
\email{takizawa@tap.st.hirosaki-u.ac.jp}
\author{Hideki Asada}
\email{asada@hirosaki-u.ac.jp}
\affiliation{
Graduate School of Science and Technology, Hirosaki University,
Aomori 036-8561, Japan}
\date{\today}

\begin{abstract}

This paper extends the de-Sitter/anti-de Sitter (dS/AdS)
background method
based on the optical metric
for gravitational lens
[Phys. Rev. D {\bf 105}, 084022 (2022)]
to a static optical constant-curvature (SOCC) background.
It is shown that
the exact lens equation on the SOCC background
can be written in the same form as that for
either Minkowski, dS or AdS background
in terms of flat, spherical or hyperbolic trigonometry,
depending on the Gaussian curvature
of the equatorial plane in the SOCC background.
To exemplify the SOCC method,
we consider the gravitational lens in Mannheim-Kazanas (MK) solution of Weyl gravity,
which includes Rindler and de Sitter terms.
In the zero mass limit,
the deflection angle of light for the MK solution in the literature
diverges to infinity.
This is because there is a self-contradiction in their perturbative approximations
of the MK metric and the orbit equation.
The SOCC method incorporates the long-distance curvature effect into the background.
Thereby the SOCC expression for the deflection angle of light in the MK solution
is finite also in the zero mass limit.
\end{abstract}


\maketitle

\section{Introduction}

The gravitational deflection of light
has attracted a lot of interest since the first experimental confirmation of
the theory of general relativity
\cite{Eddington,Will}.
In particular, the direct images of the vicinity of supermassive black hole
candidates by Event Horizon Telescope are currently  increasing the importance
of the gravitational deflection of light
\cite{EHT2019, EHT2022}.

Conventional formulations of the gravitational lens
are very useful
\cite{SEF, Petters, Dodelson, Keeton},
where
the asymptotic flatness of a spacetime is usually assumed.
However, the gravitational lens plays a role also
(1) in cosmology in the presence of the cosmological constant
and (2) asymptotically nonflat solutions in modified theories of gravity.
In order to properly investigate gravity at long distance,
the conventional method is not adequate.

The main purpose of the present paper
is to extend the de-Sitter/anti-de Sitter (dS/AdS)
background method
based on the optical metric
for gravitational lens
\cite{Takizawa2022}
to a static optical constant-curvature (SOCC) background.
We shall show that
the exact lens equation on the SOCC background
can be rewritten in the same form as that for either Minkowski,
dS or AdS background
in terms of flat, spherical or hyperbolic trigonometry
\cite{Anderson-book, Ratcliffe-book},
depending on the Gaussian curvature
of the equatorial plane in the SOCC background.

To exemplify the SOCC method,
we shall consider also Mannheim and Kazanas (MK) solution
of Weyl conformal gravity
\cite{MK}.
In the zero mass limit,
the deflection angle of light for the MK solution in the literature
\cite{Sultana, Cattani, Bhattacharya2010, Bhattacharya2011, Sultana2013}
diverges to infinity.
We shall point out that
there is a self-contradiction in perturbative approximations
of the MK metric and the orbit eqatuation.
In this regard,
the SOCC method incorporates the long-distance curvature effect into the background
in a self-consistent manner.
Thereby the SOCC expression for the deflection angle of light in the MK solution
is finite also in the zero mass limit.

This paper is organized as follows.
In Section II, we define a SOCC
background space
by using the optical metric.
Section III discusses the gravitational lens
on the SOCC background
in terms of flat, spherical or hyperbolic trigonometry.
In Section IV,
light rays in the MK solution are reexamined
in the SOCC approach.
Section V summarizes this paper.
Throughout this paper, we use the unit of $G = c = 1$.

\section{Background and lens}
\subsection{Background spacetime metric}
We examine the light deflection due to a localized lens object. We denote lens parameters as $p_i$ where $i$ labels each parameter of the lens object (e.g. mass, spin parameter, electric charge, and so on). The zero point of $p_i\,(p_i = 0)$ is chosen when the lens object does not exist. There can be global parameters other than the localized lens parameters, which are denoted by $q_i$ (e.g. the cosmological constant $\Lambda$). If $q_i \neq 0$, the background is not Minkowskian.

The spacetime metric can be written as
\begin{align}
  \label{Full-metric}
  ds^2 = g_{\mu\nu}(p_i, q_i)\,dx^{\mu} dx^{\nu}\,\,.
\end{align}
We can thus define a background metric as
\begin{align}
  \label{Def-bg}
  \bar{g}_{\mu\nu} \equiv g_{\mu\nu}(p_i, q_i) |_{p_i=0}\,\,,
\end{align}
where an overbar denotes a quantity in the background.

\subsection{Background optical metric}
In this paper, we focus on a static and spherically symmetric spacetime.
The line element can be written as
\begin{align}
  ds^2 = -A(r)dt^2 + B(r)dr^2 + C(r) d\Omega^2\,\,,
\end{align}
where $d\Omega^2 \equiv d\theta^2 + \sin^2\theta\,d\phi^2$. For this metric, we treat the optical metric followed by the null condition $ds^2 = 0$, which determines the light propagation e.g.~\cite{Ishihara2016, GW, AK2000}.
The optical metric describes a three-dimensional Riemannian space, where the distance between two points is given by $d\ell^2 \equiv dt^2$.

Since the spacetime is spherically symmetric, the photon orbital plane can be chosen as the equatorial plane without the loss of generality. Hence the hypersurface which is defined by the optical metric becomes a two-dimensional Riemannian space.
It is a hypersurface at $\theta=\pi/2$.

We can write the optical metric for this two-dimensional space as
\begin{align}
  d \ell^2 &\equiv \gamma_{IJ} dx^I dx^J \notag\\
  &= \frac{B(r)}{A(r)} dr^2 + \frac{C(r)}{A(r)} d\phi^2\,\,,
\end{align}
where subscripts denote the spatial coordinates $r$ and $\phi$.
According to the definition of the background, a background optical metric is defined as
\begin{align}
  \bar{\gamma}_{IJ} &\equiv \gamma_{IJ} |_{p_i=0}\,\,.
\end{align}
For later convenience, we define the circumference radius on the optical metric as
\begin{align}
  \tilde{r} &\equiv \sqrt{\frac{\bar{C}(r)}{\bar{A}(r)}}\,\,.
\end{align}
We define also a function which represents the inverse of the radial component of the optical metric with respect to $\tilde{r}$ as
\begin{align}
  \bar{F}(\tilde{r}) &\equiv \frac{1}{\bar{\gamma}_{\tilde{r}\tilde{r}}}\,\,.
\end{align}
Thereby, the line element of the background space can be written in terms of $\tilde{r}$ as
\begin{align}
  \label{metric-bg}
  d\bar{\ell}^2 = \frac{d\tilde{r}^2}{\bar{F}(\tilde{r})} + \tilde{r}^2 d\phi^2\,\,.
\end{align}

\section{SOCC approach}
\subsection{SOCC background}
In the above definitions of the background and the gravitational lens, the background is not necessarily flat.
In a non-flat background space, a conventional gravitational lens equation based on the Euclidean geometry is not valid. Deviations may be significant at  large distance
\cite{Takizawa2022}.

Furthermore, the present paper focuses on a background with a constant Gaussian curvature. In differential geometry, a $N$-dimensional constant-curvature space is isometric (under the assumptions of simple connectivity and completeness) to either Euclidean space $\mathbb R^N$, sphere $S^N$, or hyperbolic space $H^N$ space.

Hence a geometry of the background must be either Euclidean, spherical, or hyperbolic, depending on the sign of the constant Gaussian curvature of the background.

The Gaussian curvature for Eq.~(\ref{metric-bg}) can be calculated as \cite{Werner2012}
\begin{align}
  \bar{K}^{\mathrm{opt}} &= \frac{R_{\tilde{r}\phi \tilde{r}\phi}}{\mathrm{det}(\bar{\gamma}_{IJ})}\,\,.
\end{align}
Since the metric depends only on the radial coordinate, the condition for being the SOCC background is
\begin{align}
  \label{3rd}
  \frac{\partial \bar{K}^{\mathrm{opt}}}{\partial r} = 0\,\,.
\end{align}

We impose the regularity condition $\bar{F}(\tilde{r}) \neq 0$ and the normalization $\bar{F}(\tilde{r})|_{\tilde{r}=0} = 1$.
Then, a general solution of Eq.~(\ref{3rd}) is obtained as
\begin{align}
  \label{F(r)}
  \bar{F}(\tilde{r}) = 1 + \kappa \tilde{r}^2\,\,,
\end{align}
where $\kappa$ is an integration constant.

Therefore, the optical metric for the background space for Eq.~(\ref{F(r)}) describes a constant-curvature space \cite{CCS}.
The Gaussian curvature for Eq.~(\ref{F(r)}) is related to $\kappa$ as
\begin{align}
  \bar{K}^{\mathrm{opt}} = -\kappa\,\,.
\end{align}
Flat trigonometry based on the Euclidean geometry can be used for flat SOCC backgrounds $(\bar K^{\mathrm{opt}}=0)$, while spherical and hyperbolic trigonometry are used for positive SOCC backgrounds $(\bar K^{\mathrm{opt}}>0)$ and negative SOCC backgrounds $(\bar K^{\mathrm{opt}}<0)$, respectively.

Any metric function quadratic in $r$ can satisfy Eq.~(\ref{F(r)}),
after the origin of $r$ is appropriately shifted.
One example is de Sitter (or AdS) metric. This has been recently investigated \cite{Takizawa2022}. Another interesting example is the massless case of the Mannheim-Kazanas (MK) metric in conformal Weyl gravity \cite{MK}. This case is discussed in Section $\mathrm{IV}$.

\subsection{Exact lens equation on SOCC background}
The exact lens equation
on a flat background is derived in Reference \cite{Bozza2008}.
In a unified manner,
the exact lens equation is
extended to a spherical background and a hyperbolic one
by using trigonometry for dS/AdS background, respectively
\cite{Takizawa2022}.
Can a SOCC situation be described
by using the exact lens equation dedicated to dS/AdS?

We should observe that Eqs. (\ref{metric-bg}) and (\ref{F(r)})
for any static SOCC background
can be rewritten in the same form as dS/AdS cases.
By direct calculations,
one can show that
Eqs. (7) and (10) in Reference \cite{Takizawa2022}
for the equatorial case of the optical metric for dS/AdS
become Eq. (\ref{metric-bg}) with Eq. (\ref{F(r)}),
where $\tilde r = \sinh\rho$ and $\tilde r = \sin\rho$ are
used, respectively.
Here, dS/AdS cases
correspond to $\kappa = -1$ and $1$ in Eq. (\ref{F(r)}), respectively.
Therefore,
the exact lens equation in Reference \cite{Takizawa2022}
can be used for
a static SOCC background.
The flat, spherical or hyperbolic trigonometry for the exact lens equation
depends on the sign of the Gaussian curvature
of the equatorial plane in the SOCC background.

For the later convenience,
we briefly summarize the exact lens equation \cite{Takizawa2022}.
 In terms of angular diameter distances rescaled by $\bar K^{\mathrm{opt}}$,
the gravitational lens equation taking account of the background geometry
is found in Eq.~(72) in Reference~\cite{Takizawa2022}
\begin{align}
  \label{lenseq-all-D}
\alpha - \theta
=&
\arcsin\left(
\sqrt{
\frac{1 + \mathcal{K} \mathcal D_S^2 \tan^2\beta}{\mathcal{D}_{LS}^2 + \mathcal{D}_S^2 \tan^2\beta}
}
\mathcal D_L\sin\theta\right)
\notag\\
&
-
\arctan\left(\frac{\mathcal D_S}{\mathcal D_{LS}}\tan\beta\right)\,\,,
\end{align}
where $\beta$ is the angular position in the absence of a lens. $\theta$ and $\alpha$ denote the image position angle and the deflection angle of light, respectively~(See Figures 1-3).
Here, $D$ denotes a physical distance  in a spherical (or hyperbolic) background,
and  non-dimensional distance $\mathcal D$ is rescaled
in terms of a curvature radius of the background space,
given by
$\mathcal{D} = D  \sqrt{\bar{K}^{\mathrm{opt}}}$
(or  $\mathcal{D} = D  \sqrt{-\bar{K}^{\mathrm{opt}}}$).
Eq. (\ref{lenseq-all-D}) is unified in terms of $\mathcal{K}$
which denotes $+1$, $0$ and $-1$ for spherical, flat, hyperbolic geometry, respectively.

If Eq.~(\ref{lenseq-all-D}) is expressed with the physical distance $D$, the lens equations for $\mathcal{K} = +1$ or $-1$ take the same form as
\begin{align}
  \label{LE-Kopt}
\alpha - \theta
=&
\arcsin\left(
\sqrt{
\frac{1 + \bar K^{\mathrm{opt}} D_S^2 \tan^2\beta}{D_{LS}^2 + D_S^2 \tan^2\beta}
}
D_L\sin\theta\right)
\notag\\
&
-
\arctan\left(\frac{D_S}{D_{LS}}\tan\beta\right)\,\,.
\end{align}

For $\bar K^{\mathrm{opt}}=0$, Eq.~(\ref{LE-Kopt}) recovers the exact lens equation for the flat background
e.g.
\cite{Bozza2008, Takizawa2020b}.
The lens equation does not use the small-angle approximations, nor the conventional assumption that intersection of the tangents of light ray at a source and a receiver lies on the lens plane.

\begin{figure}[ht]
  \label{fig-1}
  \includegraphics[width=8.6cm]{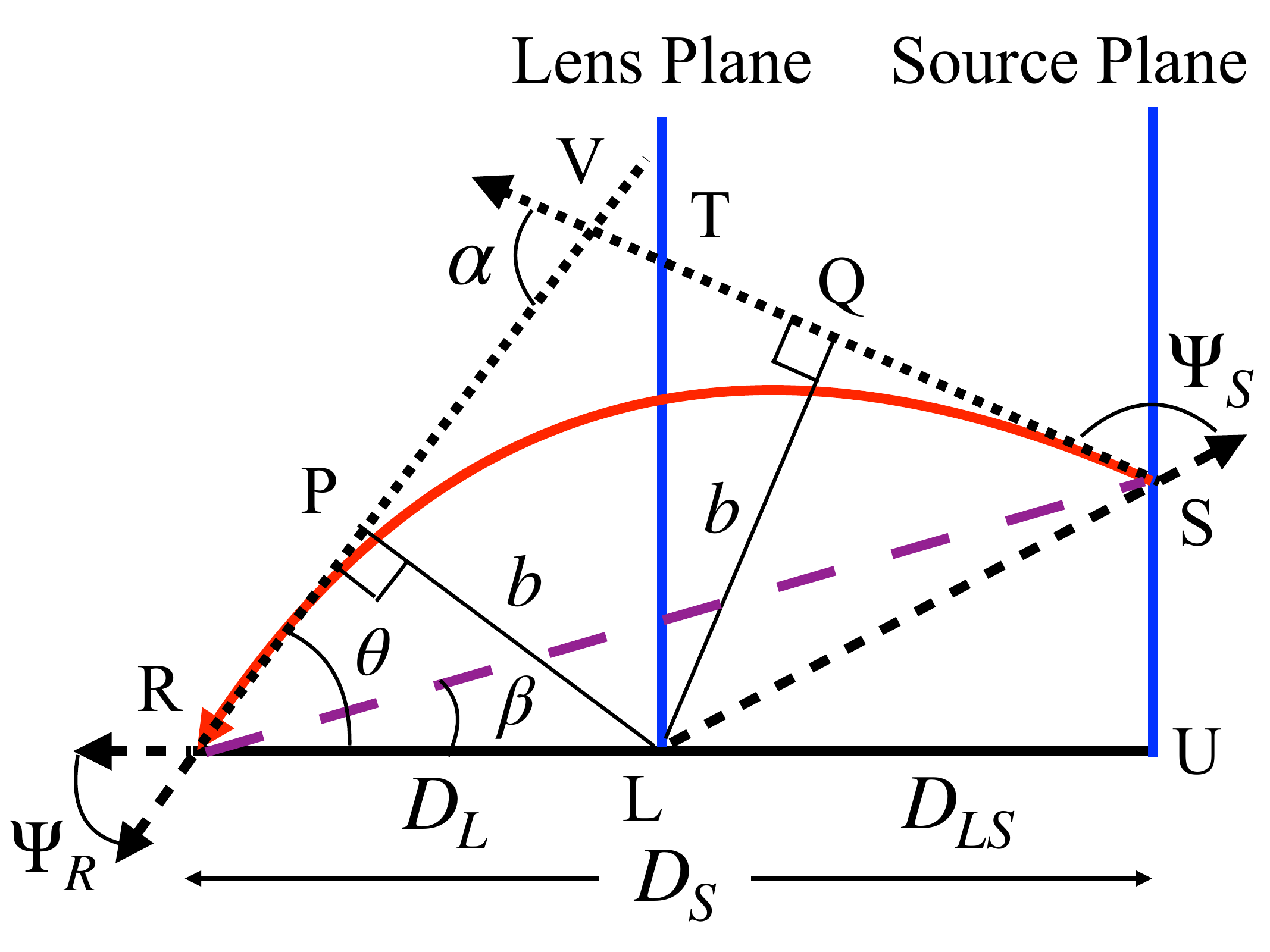}
  \caption{Schematic figure of a lens $L$, receiver $R$ and source $S$
  in a flat SOCC (Euclidean) background\cite{Takizawa2020b}.
  The red (in color) curve connecting R and S denotes a {\it lensed} light ray,
  which is not a geodesic in the Euclidean space.
  A dotted geodesic curve emanating from S is a tangent
  to the lensed light ray,
  while the other dotted one from R is another tangent to the light ray.
  The latter dotted line indicates the lensed image direction
  $\theta (= \Psi_R)$
  seen from the receiver.
  A geodesic curve between R and S is denoted by a long dashed line,
  which indicates the unlensed source direction $\beta$.
  The lens and source planes are vertical to the geodesic line RU.
  The lens and source planes are parallel to each other,
  because they live in a Euclidean space.}
\end{figure}

\begin{figure}
  \label{fig-2}
  \includegraphics[width=8.6cm]{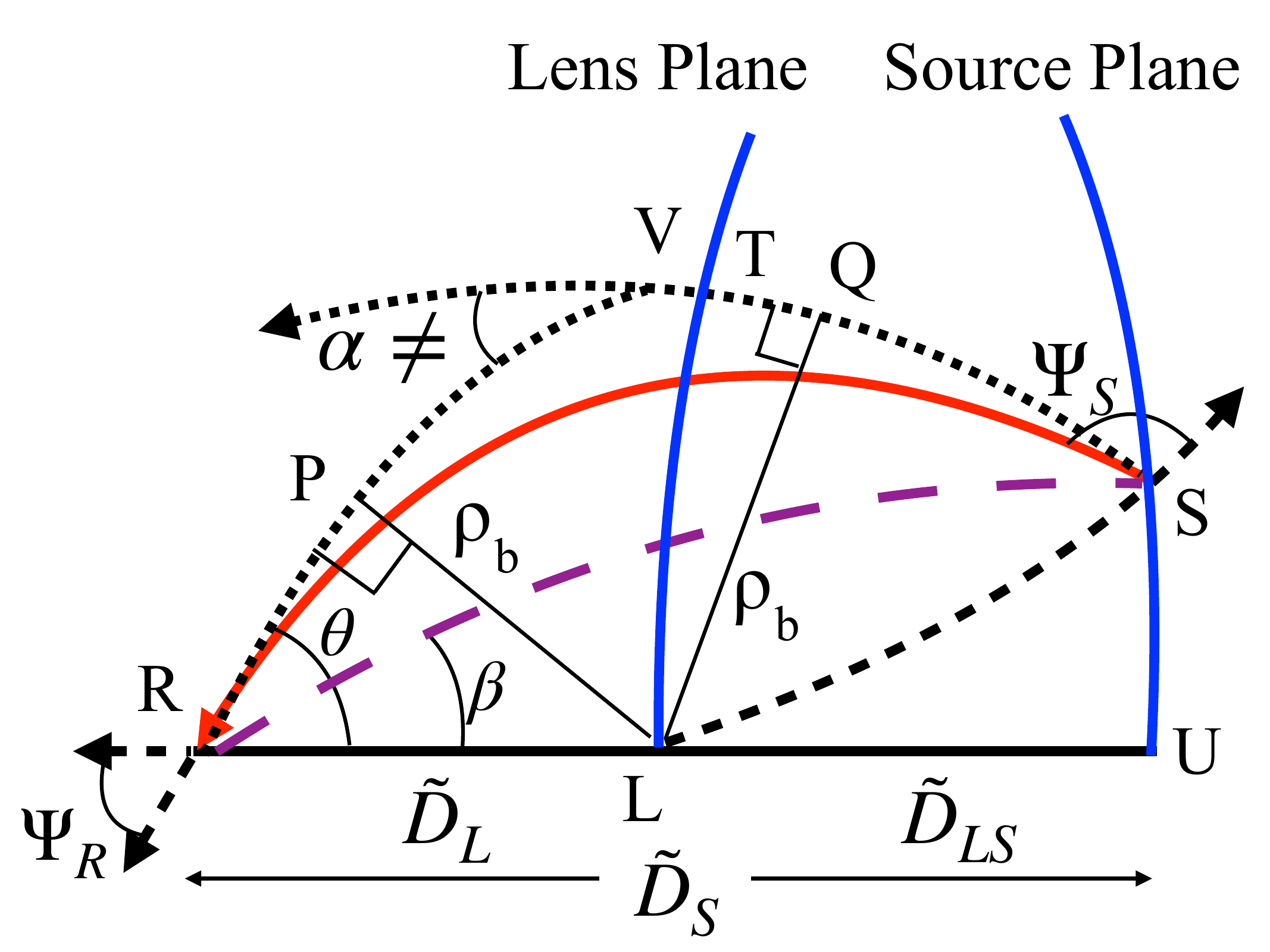}
  \caption{
  A gravitational lens configuration in a spherical background. The notations are the same as in Figure 1. For its simplicity, a line connecting L and R is drawn as a straight one
  for reference.
 The blue (in color)
  curves which denote lens plane and source one are not parallel to each other
  because of sphericity.
  Note that the sum of the inner angles for the spherical quadrilateral
  LRVS does not equal to $2\pi$ according to the Gauss-Bonnet theorem for the curved surface (See e.g.~\cite{Ishihara2016, Takizawa2020b}).
  Therefore,
  the outer angle at the intersection point V of the two tangents in the spherical plane differs from
  $\alpha$ assuming flat background. (In this sense, $\alpha \neq 0$ is described in Figure 1.)}
\end{figure}

\begin{figure}
  \label{fig-3}
  \includegraphics[width=8.6cm]{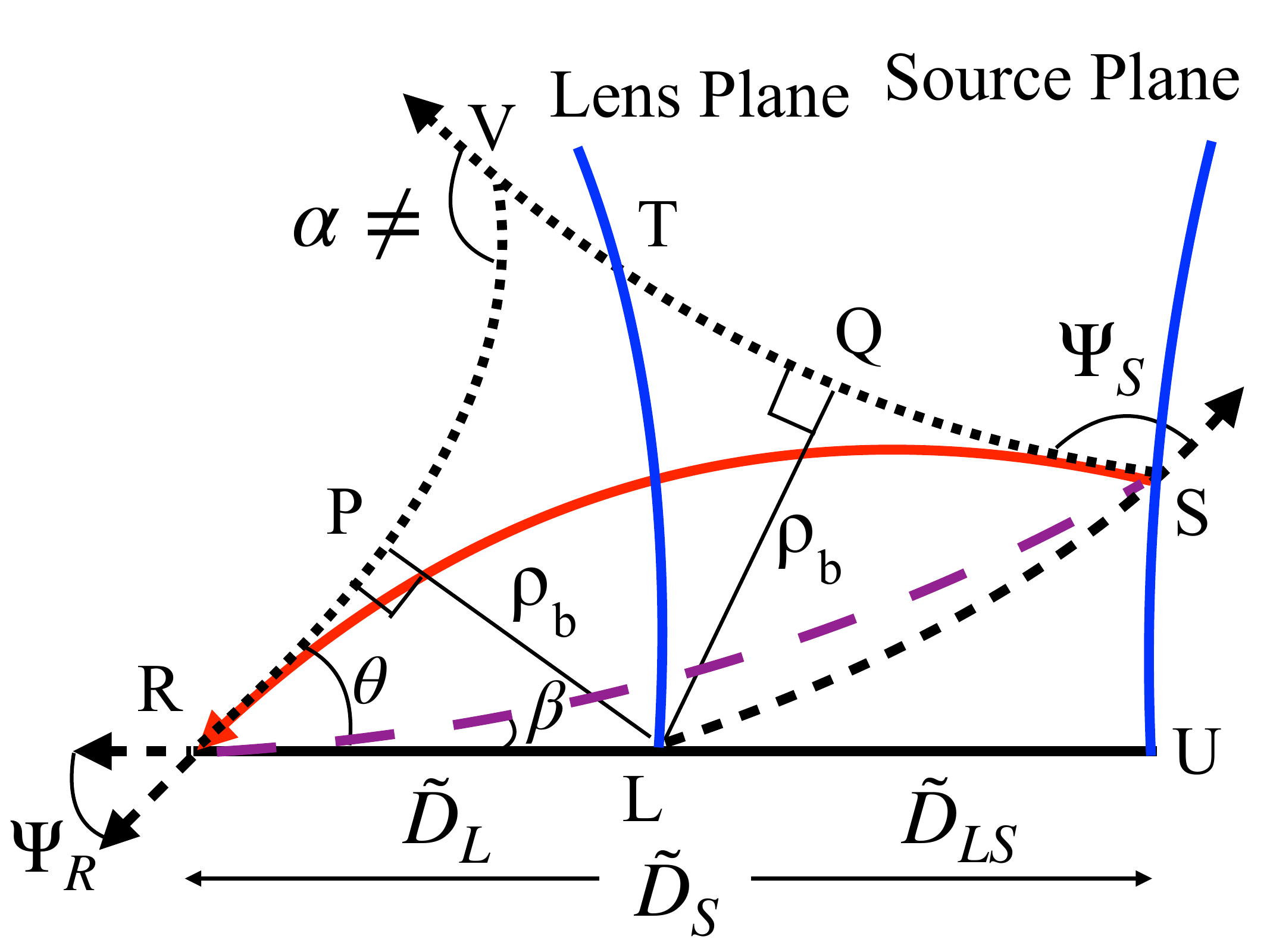}
  \caption{
  A gravitational lens configuration in a hyperbolic background.
  The blue (in color)
  curves which denote lens plane and source one are not parallel to each other
  because of hyperbolicity.
  Similarly to the configuration in spherical background, the sum of the inner angles for the hyperbolic quadrilateral
  LRVS does not equal to $2\pi$ according to the Gauss-Bonnet theorem for the curved surface.
  Therefore, the outer angle at the intersection point V of the two tangents in the hyperbolic plane differs from
  $\alpha$ assuming a flat background.}
\end{figure}

\subsection{Iterative solutions}

In order to solve the lens equation analytically,
we consider an iterative method with assuming small angles. From the relation between an impact parameter of a light ray $b$ and an image position $b = D_L \sin \theta$, the angular position of a lensed image can be approximated as $\theta \sim b / D_L$. Since this ratio is sufficiently smaller than unity under the small angle approximation, let $b / D_L$ be $\mathcal{O}(\varepsilon)$, where $\varepsilon$ is a non-dimensional small parameter.

Hence we consider a Taylor-series
expansion of $\theta$ as
\begin{align}
  \label{expand-theta}
  \theta &= \sum_{n=1}^{\infty} \varepsilon^n \theta_{(n)}\,\,,
\end{align}
where the expansion parameter $\varepsilon$ plays a role as a book-keeping.
As we shall discuss below,
if we assume $m \ll b$, the ratio $m / b$ is much smaller than unity, where $m$ denotes a lens mass. In this weak deflection approximation, therefore, $\alpha \sim m / b = \mathcal{O}(\epsilon)$. In the weak deflection and small angle approximations, we assume $\varepsilon / \epsilon = \mathcal{O}(1)$. Hence, $\alpha$ can be expanded in terms of $\varepsilon$ as
\begin{align}
  \label{expand-alpha}
  \alpha &= \sum_{n=1}^{\infty} \varepsilon^{n} \alpha_{(n)}\,\,.
\end{align}

As shown below,
$\alpha \sim m / b \sim m / D_L \theta$.
From $ m / D_L \propto \varepsilon^2$ and $\theta = \mathcal{O}(\varepsilon)$, $\alpha$ is $\mathcal{O}(\varepsilon)$ in our expansion method. $\beta$ is given by hand. We examine a case that multiple images occur, for which $\beta \sim \theta_E \sim \mathcal{O}(\varepsilon)$.
This is because $\theta_E \sim \sqrt{m / D_L} = \mathcal{O}(\varepsilon)$.
Therefore, we take
$\beta = \varepsilon \beta_{(1)}$.

The solutions of Eq.~(\ref{LE-Kopt}) are obtained up to the fourth order in $\varepsilon$ as
\begin{align}
  \label{theta-1st}
  \theta_{(1)} &= \frac{D_S}{D_L + D_{LS}}\beta_{(1)} + \frac{D_{LS}}{D_L + D_{LS}}\alpha_{(1)}\,\,,\\
  \label{theta-2nd}
  \theta_{(2)} &= \frac{D_{LS}}{D_L + D_{LS}}\alpha_{(2)}\,\,,
\end{align}
and
\begin{align}
  \label{theta-3rd}
  \theta_{(3)} = &\frac{D_{LS}}{D_L + D_{LS}} \alpha_{(3)} \notag\\
  &+ \frac{1}{3}\frac{D_S}{D_L + D_{LS}}\left[1 - \left(\frac{D_L}{D_{LS}}\right)^2\right]\beta_{(1)}^3\notag\\
  &+ \frac{1}{2}\left(\frac{D_S}{D_{LS}}\right)^2 \frac{D_L (1 - \bar K^{\mathrm{opt}} D_{LS}^2)}{D_L + D_{LS}}\beta_{(1)}^2 \theta_{(1)}\notag\\
  &+ \frac{1}{6}\frac{D_L}{D_L + D_{LS}}\left[1 - \left(\frac{D_L}{D_{LS}}\right)^2\right] \theta_{(1)}^3\,\,.
\end{align}

In order to fully consider the constant-curvature effect of the background, the background dependence of the deflection angle of light should be clarified. This clarification will be done by using a concrete example in Section $\mathrm{IV}$.

\section{Example: Gravitational lensing in MK background}
\subsection{Mannheim-Kazanas solution}
In this section, we shall show how to use the SOCC approach for the gravitational lens. We consider the MK solution in the conformal Weyl gravity \cite{MK}.
The line element is
\begin{align}
  ds^2 = -f(r) dt^2 + \frac{1}{f(r)}dr^2 + r^2 d\Omega^2 ,
\end{align}
where the metric function is given by
\begin{align}
  \label{MK-full-metric}
  f(r) \equiv 1 - 3\beta\gamma - \frac{\beta(2 - 3\beta\gamma)}{r} + \gamma r - kr^2 ,
\end{align}
and $\beta, \gamma$ and $k$ are integration constants.
We follow Reference \cite{MK}
to use $\beta$ for characterizing the MK solution
only in Eqs. (\ref{MK-full-metric}) - (\ref{1-3betagamma}),
while the rest of the present paper follows
the standard theory of gravitational lens \cite{SEF}
to use the conventional notation for the intrinsic source position as $\beta$.
This does not confuse readers.

Here, we define
\begin{align}
  \label{mass}
  2m \equiv \beta(2 - 3\beta\gamma)\,\,,
\end{align}
where $m$ denotes a mass of a lens object. From Eq.~(\ref{mass}) we obtain
\begin{align}
  \label{1-3betagamma}
  1 - 3\beta\gamma = \mp \sqrt{1 - 6m\gamma}\,\,.
\end{align}
We assume $1 - 3\beta\gamma > 0$. Then, the positive sign is adopted in Eq.~(\ref{1-3betagamma}). Hence we can reexpress Eq.~(\ref{MK-full-metric}) as
\begin{align}
  \label{MK-m}
  f(r) = \sqrt{1 - 6m\gamma} - \frac{2m}{r} + \gamma r - k r^2\,\,.
\end{align}
This form of $f(r)$ agrees with that in Reference~\cite{Cattani}.

\subsection{Metric splitting}
We have discussed how to split a full metric into a background part and a lens part
(as a perturbation)
in Section $\mathrm{II}$. By using Eqs.~(\ref{Full-metric}) and (\ref{Def-bg}), let us split the metric function as
\begin{align}
  f(r) = f(r)_{\mathrm{bg}} + f(r)_{\mathrm{Lens}}\,\,.
\end{align}
In Eq.~(\ref{MK-m}), however, we cannot split the metric function in a straightforward manner because of the existence of $\sqrt{1 - 6 m \gamma}$. For this reason, we shall perform the following coordinate and conformal transformations.

First, we consider the coordinate transformation as
\begin{align}
  x^{\mu} = (t, r, \theta, \phi) \longrightarrow \hat{x}^{\mu} = (\hat{t}, r, \hat{\theta}, \hat{\phi})\,\,,
\end{align}
where the new coordinates with the hat are
\begin{align}
  \hat{t} = a\,t\,\,,\,\,\hat{\theta} = \sqrt{a}\,\theta\,\,,\,\,\hat{\phi} = \sqrt{a}\,\phi ,
\end{align}
and $a$ denotes the constant term as
\begin{align}
  a \equiv \sqrt{1 - 6m\gamma}\,\,.
\end{align}
Therefore, the MK metric in the coordinate system $\hat{x}^{\mu}$ is rewritten as
\begin{align}
  \label{MK-hat}
  ds^2 &= g_{\mu\nu}(\hat{x}) d\hat{x}^{\mu} d\hat{x}^{\nu}\notag\\
  &= -\frac{f(r)}{a^2} d\hat{t}^2 + \frac{1}{f(r)}dr^2 + \frac{r^2}{a}d\hat{\Omega}^2\,\,,
\end{align}
where $d\hat{\Omega}^2 \equiv d\hat{\theta}^2 + \sin^2 (\hat{\theta}/\sqrt{a})d\hat{\phi}^2$\,\,.

Next, we consider the following conformal transformation as
\begin{align}
  \hat{g}_{\mu\nu}(\hat{x}) = a\,g_{\mu\nu}(\hat{x})\,\,.
\end{align}
After these transformations, we obtain
\begin{align}
  \label{MK-rescaled}
  d\hat{s}^2 &\equiv a\,ds^2\notag\\
  &= -\mathcal{F}(r)d\hat{t}^2 + \frac{1}{\mathcal{F}(r)}dr^2 + r^2 d\hat{\Omega}^2\,\,,
\end{align}
where the metric function is
\begin{align}
  \mathcal{F}(r) &\equiv \frac{f(r)}{a} \notag\\
  &= 1 - \frac{2 \hat{m}}{r} + \hat{\gamma}r - \hat{k}r^2\,\,,
\end{align}
and parameters are scaled as
\begin{align}
  \hat{m} \equiv \frac{m}{a}\,\,,\,\,\hat{\gamma} \equiv \frac{\gamma}{a}\,\,,\,\,
  \hat{k} \equiv \frac{k}{a}\,\,.
\end{align}

For $d\hat{s}^2$, we can easily split the metric into the background and the lens part as
\begin{align}
  \mathcal{F}(r) = \mathcal{F}(r)_{\mathrm{bg}} + \mathcal{F}(r)_{\mathrm{lens}}\,\,,
\end{align}
where
\begin{align}
  \label{MK-bg}
  \mathcal{F}(r)_{\mathrm{bg}} &= 1 + \hat{\gamma} r - \hat{k} r^2\,\,,\\
  \label{MK-lens}
  \mathcal{F}(r)_{\mathrm{lens}} &= -\frac{2\hat{m}}{r}\,\,.
\end{align}

Henceforth, we focus on a case that the receiver and source are located outside a horizon if the horizon exists. In order to avoid a light ray crossing a horizon in the background spacetime, we focus on the range of $r$ where $\mathcal{F}(r)_{\mathrm{bg}} > 0$.
We consider all cases of the sign of the parameters $\hat\gamma$ and $\hat k$ separately.

\noindent$(\bm{\mathrm{i}) \,\,\hat{k}>0 \,\,\mathrm{case}}$. \\
From $\mathcal{F}(r)_{\mathrm{bg}} > 0$, $r \geq 0$ and Eq.~(\ref{MK-m}),
we obtain
\begin{align}
  \label{r_range_k+}
   r < \frac{\hat\gamma}{2\hat k} + \sqrt{\frac{\hat{\gamma}^2 + 4\hat k}{4\hat{k}^2}}\,\, ,
\end{align}
where
the horizon exists at
\begin{align}
  \label{+horizon}
  r_H = \frac{\hat\gamma}{2\hat k} + \sqrt{\frac{\hat{\gamma}^2 + 4\hat k}{4\hat{k}^2}}\,\,.
\end{align}
We refer to $r_H$ as MK horizon.

Regardless of the sign of $\hat\gamma$, $r_H \geq 0$ is satisfied; even if $\hat\gamma < 0$, the absolute value of the first term in $r_H$ is smaller than the second term, which leads to $r \geq 0$.

\noindent$(\bm{\mathrm{ii}) \,\,\hat{k}<0 \,\,\mathrm{case}}$.\\
If $\hat\gamma < 0$ and $\hat{\gamma}^2 + 4\hat{k} < 0$,
$r$ can be positive,  where
this spacetime has no horizon.
If $\hat\gamma < 0$ and $\hat{\gamma}^2 + 4\hat{k} > 0$, the necessary and sufficient condition that a positive $r$ satisfies $\mathcal{F}(r)_{\mathrm{bg}} > 0$ is
 \begin{align}
   \label{r_range_k-}
    r < \frac{\hat\gamma}{2\hat k} - \sqrt{\frac{\hat{\gamma}^2 + 4\hat k}{4\hat{k}^2}}\,\,.
 \end{align}
The MK horizon is thus located at
\begin{align}
  \label{-horizon}
  r_H = \frac{\hat\gamma}{2 \hat k} - \sqrt{\frac{\hat{\gamma}^2 + 4\hat k}{4\hat{k}^2}}\,\,.
\end{align}

If $\hat{\gamma} > 0$, on the other hand, any $r$ means a location
inside the horizon.
Hence $\hat{k} < 0$ and $\hat{\gamma} > 0$ are not
considered in this paper.

\noindent$(\bm{\mathrm{iii}) \,\, \hat{k}=0 \,\,\mathrm{case}}$.\\
For $\hat{\gamma} > 0$ to satisfy the condition $\mathcal{F}(r)_{\mathrm{bg}} > 0$, $r > - 1 / \hat\gamma$.
Since $r \geq 0$, $r$ always satisfies this condition. For the positive $\hat\gamma$, the horizon does not exist, whereas for $\hat{\gamma} < 0$, MK horizon is located at $r_H = 1 / |\hat{\gamma}|$.

We should note that the metric for $\hat{\gamma} = 0$ corresponds to the dS/AdS metric for the static slicing, where $\hat k$ can be interpreted as the cosmological constant.

\subsection{SOCC approach for MK lensing}
Eq.~(\ref{MK-bg}) includes the linear-$r$ term (called Rindler term) and $r^2$ term (called de Sitter term). In this paper, we refer to Eq.~(\ref{MK-bg}) as
 MK background. For the background metric, we can obtain the optical metric as
\begin{align}
  \label{MK_opt}
  d \bar{\ell}^2 = [\mathcal{F}(r)_{\mathrm{bg}}]^{-2} dr^2 + r^2 [\mathcal{F}(r)_{\mathrm{bg}}]^{-1} d\hat\phi^2\,\,.
\end{align}
The metric function for $\hat k \neq 0$ can be rewritten as
\begin{align}
  \mathcal{F}(r)_{\mathrm{bg}} = -\frac{\hat K^{\mathrm{opt}}}{\hat{k}} \left[1 + \frac{\hat{k}^2}{\hat{K}^{\mathrm{opt}}}\left(r - \frac{\hat{\gamma}}{2\hat{k}}\right)^2 \right]\,\,,
\end{align}
where $\hat{K}^{\mathrm{opt}}$ denotes the Gaussian curvature which is calculated by using the background optical metric as
\begin{align}
  \label{K_opt_bg}
  \hat{K}^{\mathrm{opt}} = - \frac{\hat{\gamma}^2}{4} - \hat{k}\,\,.
\end{align}

Eq.~(\ref{K_opt_bg}) indicates that the hypersurface described by the optical metric in the MK background has a constant curvature which is parameterized by $\hat{k}$ and $\hat{\gamma}$. The sign of $\hat{\gamma}$ does not affect the value of $\hat{K}^{\mathrm{opt}}$.
Therefore, $\hat{K}^{\mathrm{opt}}$ for $\hat{k} = 0$ space is always negative for $\hat{\gamma} \neq 0$; the space is hyperbolic regardless of the sign of $\hat{\gamma}$. Even if $\hat{k}$ is positive, the space is still hyperbolic.
However, we should
note that, if $\hat{k}$ is negative, then $\hat{K}^{\mathrm{opt}}$ can take any sign. In this case, the background geometry is to be either hyperbolic or spherical depending on $\hat{k}$ and $\hat{\gamma}^2 / 4$.

The background geometry is either Euclidean, hyperbolic or spherical in the Riemannian geometry.
In the following, we shall introduce useful coordinates to clarify the background geometry depending on $\hat{k}$ and $\hat{\gamma}$.

\noindent$(\bm{\mathrm{i}) \,\,\hat{k}>0 \,\,\mathrm{case}}$. \\
For later convenience, we introduce a new radial coordinate as
\begin{align}
  \label{R-k+}
  R \equiv \sqrt{\frac{\hat{k}^2}{-\hat{K}^{\mathrm{opt}}}}\,\left(r - \frac{\hat{\gamma}}{2\hat{k}}\right)\,\,,
\end{align}
such that the optical metric in Eq.~(\ref{MK_opt}) can be conformally rescaled as
\begin{align}
  \label{dl_hat_k+}
  d\hat{\ell}^2 &\equiv -\hat{K}^{\mathrm{opt}} d\bar \ell^2 \notag\\
  &= \left(\frac{dR}{1 - R^2}\right)^2 - \frac{\hat{K}^{\mathrm{opt}}\left(R + \cfrac{\hat \gamma}{2\sqrt{-\hat{K}^{\mathrm{opt}}}}\right)^2}{\hat{k}(1 - R^2)}d{\hat\phi}^2\,\,.
\end{align}
By performing another coordinate transformation
\begin{align}
  \label{R-k+2}
  R \equiv \tanh\eta\,\,,
\end{align}
and denoting a constant in the second term of Eq.~(\ref{dl_hat_k+}) as
\begin{align}
  \tanh{C_1} = \frac{\hat \gamma}{2\sqrt{-\hat{K}^{\mathrm{opt}}}}\,\,,
\end{align}
Eq.~(\ref{dl_hat_k+}) can be simplified as
\begin{align}
  \label{H-metric}
  d\hat{\ell}^2 = d\rho^2 + \sinh^2 \rho\,d\hat\phi^2\,\,,
\end{align}
where the coordinate $\rho$ is defined as $\rho \equiv \eta + C_1$.

Note that $\rho$ coordinate is different from $\chi$ coordinate that is usually used in general relativistic cosmology~\cite{footnote-1}.
Eq.~(\ref{H-metric}) clearly means that the optical metric for MK background space for $\hat{k} > 0$ describes a hyperbolic space, in which unlensed light rays are geodesic curves.
From Eqs.~(\ref{r_range_k+}) and (\ref{R-k+}), the range of $R$ coordinate is $-\hat{\gamma} / 2 \sqrt{-K^{\mathrm{opt}}} \leq R < 1$.
By using the relation between the coordinates $R$ and $\rho$, we can find $0 \leq \rho < +\infty$. Hence the $\rho$ coordinate covers the whole hyperbolic surface.

We stress that the MK horizon is located at $\rho = +\infty$ in the rescaled metric Eq.~(\ref{H-metric}). This means that the use of $\rho$ coordinate allows to easily avoid the MK horizon crossing.

\noindent$(\bm{\mathrm{ii}) \,\,\hat{k}<0 \,\,\mathrm{case}}$.\\
 Also $\hat{\gamma} < 0$ and $\hat{K}^{\mathrm{opt}} < 0$ lead to the hyperbolic space. For this case, Eq.~(\ref{R-k+}) can be used.

However, we have to do a coordinate transformation different from $\hat{k} > 0$ case, since $1 < \tanh C_1 \leq 1$ contradicts with $\hat{k} < 0$.
Instead of Eq.~(\ref{R-k+2}), let us introduce another radial coordinate as
\begin{align}
  R \equiv \coth \eta\,\,,
\end{align}
with denoting a constant term in the optical metric as
\begin{align}
  \coth C_2 =
  \frac{\hat \gamma}{2\sqrt{-\hat{K}^{\mathrm{opt}}}}\,\,.
\end{align}
The optical metric in the coordinate $\rho \equiv \eta - C_2$ is written in the same form as Eq.~(\ref{H-metric}).
From Eqs.~(\ref{r_range_k-}) and (\ref{R-k+}), the range of $R$ is to be $\hat{\gamma} / 2 \sqrt{-\hat{K}^{\mathrm{opt}}} \leq R < 1$. This corresponds to $0 \leq \rho < +\infty$. Therefore, also in this case, $\rho$ covers the whole hyperbolic surface.

On the other hand, for $\hat{K}^{\mathrm{opt}} > 0$, namely the spherical geometry, a new radial coordinate is defined as
\begin{align}
  R \equiv \sqrt{\frac{\hat{k}^2}{\hat{K}^{\mathrm{opt}}}}\,
  \left(r - \frac{\hat{\gamma}}{2\hat{k}}\right)\,\,,
\end{align}
for which the rescaled optical metric is
\begin{align}
  \label{dl_hat_k-}
  d\hat{\ell}^2  &\equiv \hat{K}^{\mathrm{opt}} d\bar \ell^2 \notag\\
  &= \left(\frac{dR}{1 + R^2}\right)^2 -  \frac{\hat{K}^{\mathrm{opt}}\left(R + \cfrac{\gamma}{2\sqrt{\hat{K}^{\mathrm{opt}}}}\right)^2}{\hat{k}(1 + R^2)}d\hat\phi^2\,\,.
\end{align}

We define a new coordinate $\eta$ by
\begin{align}
  R \equiv \tan\eta\,\,,
\end{align}
and we define a constant
\begin{align}
  \tan C_3 = \frac{\hat{\gamma}}{2\sqrt{\hat{K}^{\mathrm{opt}}}}\,\,.
\end{align}
Consequently, by defining a coordinate as $\rho \equiv \eta - C_3$, the rescaled optical metric is obtained as
\begin{align}
  \label{S-metric}
  d\hat{\ell}^2 = d\rho^2 + \sin^2 \rho\,d\hat\phi^2\,\,,
\end{align}
which is the metric of the spherical geometry.

From the above discussions, the optical metric in MK background can describe the spherical space only in $\hat{k} < 0$, $\hat{\gamma} < 0$ and $\hat{K}^{\mathrm{opt}} > 0$. Note that $r$ can take any positive value and the MK horizon does not exist for this case. It turns out that $\rho$ covers
the whole sphere.

\noindent$(\bm{\mathrm{iii}) \,\,\hat{k}=0 \,\,\mathrm{case}}$.\\
The background metric becomes
\begin{align}
  \mathcal{F}(r)_{\mathrm{bg}} = 1 + \hat{\gamma} r\,\,.
\end{align}
We normalize the radial coordinate as
\begin{align}
  R \equiv |\hat{\gamma}| \,r\,\,.
\end{align}

The conformally rescaled metric is obtained with the factor $1 / \hat{\gamma}^2$ as
\begin{align}
  \label{H-metric_k0a}
  d\hat{\ell}^2 = \left(\frac{dR}{1 + R}\right)^2 + \frac{R^2}{1 + R} d\hat\phi^2\,\,,
\end{align}
where we define a new radial coordinate as
\begin{align}
  R &\equiv e^{\eta} - 1\,\,,\\
  R &\equiv 1 - e^{-\eta}\,\,,
\end{align}
for $\hat{\gamma} > 0$ and $\hat{\gamma} < 0$, respectively.

These coordinates lead to
the same term $2(\cosh\eta - 1)$ in the second term of Eq.~(\ref{H-metric_k0a}), and we thus find the metric
\begin{align}
  d\hat{\ell}^2 &\equiv -\hat{K}^{\mathrm{opt}}|_{k=0}\,d\bar  \ell^2 \notag\\
  &= d\rho^2 + \sinh^2 \rho\,d\hat\phi^2\,\,,
\end{align}
where we define the coordinate $\rho \equiv \eta / 2$.
Note that $\hat{\gamma} > 0$ space in the coordinate $\rho$ describes the whole hyperbolic space, where the MK horizon location is $\rho \rightarrow \infty$.
Even in this case, therefore,
any light ray under consideration never crosses the horizon.
See also Table \ref{table-1} for the allowed region of $r$
and its realtion to the signs of $\gamma$, $k$ and $\hat K^{\mathrm{opt}}$.

\begin{table}[t]
  \centering
    \caption
    Background geometries depending on the sign of the parameter.
    "A" and "N/A" mean availabe  and not available respectively. Note that $\hat k > 0$ and $\hat K^{\mathrm{opt}} > 0 $ do not simultaneously occur  because $\hat K^{\mathrm{opt}} = -\hat k -\hat\gamma^2 / 4 $.
    If $\hat\gamma > 0$, $\hat k < 0$ and $\hat K^{\mathrm{opt}} < 0$, any positive $r$ is located inside the horizon.
    \vspace{1mm}
    \label{table}
    \begin{tabular}{p{4.5em} p{4.5em} p{4.5em} p{4.5em} p{4.5em}}
      \hline
      $\gamma$ & $k$ & $\hat{K}^{\mathrm{opt}}$ & $r > 0$ & \,\,\,\,$r_{H}$ \\
      \hline \hline
      $+$ & $+$ & $+$ & N/A & N/A \\
      $+$ & $+$ & $-$ & \,\,\,A & Eq.~(\ref{+horizon}) \\
      \hline
      $+$ & $-$ & $+$ & \,\,\,A & N/A \\
      $+$ & $-$ & $-$ & N/A & N/A \\
      \hline
      $-$ & $+$ & $+$ & N/A & N/A \\
      $-$ & $+$ & $-$ & \,\,\,A & Eq.~(\ref{+horizon}) \\
      \hline
      $-$ & $-$ & $+$ & \,\,\,A & N/A \\
      $-$ & $-$ & $-$ & \,\,\,A & Eq.~(\ref{-horizon}) \\
      \hline
    \end{tabular}
\label{table-1}
  \end{table}

\subsection{Deflection angle of light}
Here we follow Reference \cite{Ishihara2016} to adopt the definition of
the deflection angle of light in the optical metric.
The definition is valid even in an asymptotically non-flat spacetime
\cite{Takizawa2020a}
\begin{align}
  \label{alpha}
  \alpha \equiv \Psi_R - \Psi_S + \phi_{RS}\,\,,
\end{align}
where $\Psi_R$ and $\Psi_S$ denote the angle between the unit vector along the radial direction from the center of the lens object and a unit tangential vector along the light ray at a receiver and a source, respectively.
$\phi_{RS}$ denotes the coordinate separation angle between a source and a receiver.
Note that Reference \cite{Takizawa2020a} focuses only on the flat background case.
The present paper investigates a more general background with a constant curvature.

Thus, $\alpha$ is a function of the parameters $p_i$ and $q_j$.
\begin{align}
  \alpha = \alpha(p_i, q_j)\,\,.
\end{align}
We wish to consider only the contributions due to the lens objects in a consistent manner with the SOCC background. $\alpha(p_i=0, q_j)$ is the contribution purely from the background and  it should be subtracted from $\alpha(p_i, q_j)$. Namely, the deflection angle in the SOCC approach
should be
\begin{align}
  \label{alpha-lens}
  \alpha(p_i) \equiv \alpha(p_i, q_j) - \alpha(p_i=0, q_j)\,\,.
\end{align}

In a simple case that the metric has only the cosmological constant as the background parameter, Eq.~(\ref{alpha-lens}) recovers Eq.~(77) in Reference~\cite{Takizawa2022}.

To obtain the expression for $\phi_{RS}$, we discuss null geodesics in the rescaled MK metric Eq.~(\ref{MK-rescaled}). The constants of motion of light rays are found by noting Killing vector fields in $\hat{t}$ and $\hat{\phi}$ directions as
\begin{align}
  \hat E &= \mathcal{F}(r)_{\mathrm{bg}}\frac{d\hat{t}}{d\hat{\lambda}}\,\,,\\
  \hat J &= r^2 \frac{d\hat{\phi}}{d\hat{\lambda}}\,\,,
\end{align}
where $\hat{\lambda}$ is an affine parameter in the coordinate system $\hat{x}^{\mu}$. Then, the impact parameter of light is defined by
\begin{align}
  \hat{b} \equiv \frac{\hat J}{\hat E} = \frac{r^2}{\mathcal{F}(r)_{\mathrm{bg}}}\frac{d\hat{\phi}}{d\hat{t}}\,\,.
\end{align}

From the null condition, the orbit equation for light rays are obtained in terms of $u \equiv 1/r$ as
\begin{align}
  \label{OE-u}
  \left(\frac{d u}{d \hat{\phi}}\right)^2 = \frac{1}{\hat{b}^2}
  - u^2 - \hat{\gamma} u + \hat{k} + 2\hat{m} u^3\,\,.
\end{align}
Eq.~(\ref{OE-u}) is rewritten under the transformation $U \equiv u + \hat{\gamma} /2$ as
\begin{align}
  \label{OE-U}
  \left(\frac{d U}{d\hat{\phi}}\right)^2 = \frac{1}{\hat{B}^2} - U^2 + 2\hat{m}\left(U - \frac{\hat{\gamma}}{2}\right)^3\,\,,
\end{align}
where the Gaussian curvature term can be absorbed as
\begin{align}
  \label{B}
  \frac{1}{\hat{B}^2} = \frac{1}{\hat{b}^2} - \hat{K}^{\mathrm{opt}}\,\,.
\end{align}

Note that the negative $\hat{\gamma}$ contributes to a photon sphere, where the radius is $r_{\mathrm{PS}} = -2/\hat{\gamma}$. Thus the transformation means $U = u - u_{\mathrm{PS}}$, where $u_{\mathrm PS} \equiv 1 / r_{\mathrm PS}$. Since $U > 0$, this transformation implies for the light ray following Eq.~(\ref{OE-U}), $r < r_{\mathrm{PS}}$.

Since the $\hat{m}$-independent terms in the orbit equation are derived from
$\mathcal{F}(r)_{\mathrm{bg}}$,
we substitute the iterative solution
\begin{align}
  \label{SolutionForm}
  U(\hat{\phi}) = U_{\mathrm{bg}} + \hat{m}U_{\hat{m}} + \mathcal{O}(\hat{m}^2)\,\,,
\end{align}
into Eq.~(\ref{OE-U}). By direct calculations, the first term (background part) in Eq.~(\ref{SolutionForm}) is obtained as
\begin{align}
  U_{\mathrm{bg}} = \frac{\sin(\hat{\phi} + C_{\mathrm{bg}})}{\hat{B}}\,\,,
\end{align}
where $C_{\mathrm{bg}}$ is an integration constant.

Since the spacetime is spherically symmetric,
the light ray has reflection symmetry with respect to angular coordinate value
$\hat{\phi} = \pi \sqrt{a} / 2$
that corresponds to the closest approach. Therefore, the integration constant is determined as
\begin{align}
  C_{\mathrm{bg}} = \frac{\pi}{2}(1 - \sqrt{a})\,\,.
\end{align}
Hence the zeroth-order solution is
\begin{align}
  \label{0th}
  U_{\mathrm{bg}} &= \frac{\sin\hat{\varphi}}{\hat{B}} = \frac{\sqrt{1 - \hat{b^2}\hat{K}^{\mathrm{opt}}}}{\hat{b}}\sin\hat{\varphi}\,\,,
\end{align}
where the argument of the sine function is $\hat{\varphi} \equiv \hat{\phi} + C_{\mathrm{bg}}$. The transformations $u \longrightarrow U$ and $\hat{\phi} \longrightarrow \hat{\varphi}$ do not change the orbit equation nor its solution because $dU = du$ and $d\hat{\phi} = d\hat{\varphi}$, respectively.
Note that the square root in Eq.~(\ref{0th}) for $\hat{K}^{\mathrm{opt}} > 0$ leads to the upper bound on the impact parameter as
\begin{align}
  \label{upperbound}
  \hat{b} < \frac{1}{\sqrt{\hat{K}^{\mathrm{opt}}}}\,\,.
\end{align}
Namely, in the spherical SOCC background, the light rays with the impact parameter greater than the SOCC curvature radius are not allowed for $r < r_{\mathrm{PS}}$. Indeed, from Eq.~(\ref{OE-u}) excluding $\hat m$ term, it can be confirmed that $\hat{b} = 1 / \sqrt{\hat{K}^{\mathrm{opt}}}$ when $r = r_{\mathrm{PS}}$.

By using Eq.~(\ref{0th}), we obtain the linear term in either $\hat{m}$ or $\hat{\gamma}$ of Eq.~(\ref{SolutionForm}) as
\begin{align}
  \label{1st}
  U_{\hat{m}} = &\frac{1}{\hat{B}^2}(1 + \cos^2\hat{\varphi})\notag\\
  &- \frac{3\hat{\gamma}}{2\hat{B}}\left[\sin\hat{\varphi}
  - \left(\hat{\varphi} - \frac{\pi}{2}\right)\cos\hat{\varphi}\right]\,\,.
\end{align}

Therefore, the iterative solution to the orbit equation in terms of $1 / \hat{B}$ is
\begin{align}
  \label{U-sol}
  U(\hat{\varphi}) = &\frac{\sin\hat{\varphi}}{\hat{B}} + \frac{\hat{m}}{\hat{B}^2}(1 + \cos^2\hat{\varphi})\notag\\
  &- \frac{3\hat{m}\hat{\gamma}}{2\hat{B}}\left[\sin\hat{\varphi}
  - \left(\hat{\varphi} - \frac{\pi}{2}\right)\cos\hat{\varphi}\right]\,\, + \mathcal{O}(\hat{m}^2).
\end{align}

By using Eq.~(\ref{U-sol}), we obtain the angle as
\begin{align}
  \label{phi-sol}
  \hat{\varphi}(U) = &\arcsin(\hat{B}U) - \frac{\hat{m}(2-\hat{B}^2 U)}{\hat{B}\sqrt{1-\hat{B}^2 U^2}}\notag\\
  &-\frac{3\hat{m}\hat{\gamma}\hat{B}U}{2\sqrt{1-\hat{B}^2 U^2}} - \frac{3}{2}\hat{m}\hat{\gamma}\left[\arcsin(\hat{B}U)-\frac{\pi}{2}\right]\,\,\notag\\ &+ \mathcal{O}(\hat{m}^2).
\end{align}
Therefore, the separation angle in $\hat{\varphi}$ is
\begin{align}
  \label{phi_RS}
  \hat{\varphi}_{RS} &= \hat{\varphi}_R - \hat{\varphi}_S \notag\\
  &= \pi - \hat{\varphi}(U_R) - \hat{\varphi}(U_S)\,\,.
\end{align}

Next, let us calculate the $\Psi$ part in Eq.~(\ref{alpha}). From the definition,
\begin{align}
  \Psi = \arcsin\left[\hat{b}u \sqrt{\mathcal{F}(u)}\right]\,\,.
\end{align}
We can obtain the expression for $\Psi$ by using $\delta \equiv \mathcal{F}(u)_{\mathrm{lens}}/\mathcal{F}(u)_{\mathrm{bg}} \ll 1$ as
\begin{align}
  \label{Psi-u}
  \Psi = \arcsin\left[\hat{b}u \sqrt{\mathcal{F}(u)_{\mathrm{bg}}} \left(1 + \frac{\delta}{2}\right)\right] + \mathcal{O}\left(\delta^2\right)\,\,.
\end{align}
In Eq.~(\ref{Psi-u}), we can use the relation between $u$ and $U$ as
\begin{align}
  u \sqrt{\mathcal{F}(u)_{\mathrm{bg}}} = \sqrt{U^2 + \hat{K}^{\mathrm{opt}}}\,\,.
\end{align}
The expression for $\Psi$ in terms of $U$ is thus obtained as
\begin{align}
  \Psi(U) = &\arcsin\left(\hat{b}U \sqrt{1 + \frac{\hat{K}^{\mathrm{opt}}}{U^2}}\right)\notag\\
  &- \frac{\hat{m}\hat{b}U^2}{\sqrt{1 - \hat{b}^2 U^2 + \hat{b}^2 \hat{K}^{\mathrm{opt}}}}\left(1 - \frac{\hat{\gamma}}{2U}\right)^3 \left(1 + \frac{\hat{K}}{U^2}\right)^{-\frac{1}{2}}\notag\\
  &+\mathcal{O}(\hat{m}^2)\,\,.
\end{align}
Hence the $\Psi$ part is obtained as
\begin{align}
  \Psi_R - \Psi_S = \Psi(U_R) + \Psi(U_S) - \pi\,\,.
\end{align}

\subsection{Weak deflection of light by a small black hole}

In order to clarify which parameter contributes perturbatively to the deflection angle of light, we assume the weak field of a small black hole as $|\hat{b}^2 \hat{K}^{\mathrm{opt}}| \ll 1$ and $|\hat K^{\mathrm{opt}} / U^2| \ll 1$. Therefore, the deflection angle in the rescaled MK metric is expressed as
\begin{align}
  \label{MK-alpha}
  \alpha_{\mathrm{MK}} = &\frac{\hat{b}\hat{K}^{\mathrm{opt}}}{2} \left(\frac{\sqrt{1 - \hat{b}^2 U_S^2}}{U_S} + \frac{\sqrt{1 - \hat{b}^2 U_R^2}}{U_R}\right)\notag\\
  &-\frac{\hat{m}\hat{b}\hat{K}^{\mathrm{opt}}}{2}\left(\frac{1}{\sqrt{1 - \hat{b}^2 U_S^2}} + \frac{1}{\sqrt{1 - \hat{b}^2 U_R^2}}\right)\notag\\
  &+3\hat m \hat\gamma\left(\frac{\hat b U_S}{\sqrt{1 - \hat b^2 U_S^2}} + \frac{\hat b U_R}{\sqrt{1 - \hat b^2 U_R^2}}\right)\notag\\
  &- \frac{3}{2}\hat{m}\hat{\gamma}\left[\arcsin(\hat{b}U_S) + \arcsin(\hat{b}U_R) - \pi\right]\notag\\
  &+ \frac{2\hat{m}}{\hat{b}}\left(\sqrt{1 - \hat{b}^2 U_S^2} + \sqrt{1 - \hat{b}^2 U_R^2}\right)+\mathcal{O}(\hat{m}^2)\,\,.
\end{align}
The first term (derived from $\mathcal{F}(r)_{\mathrm{bg}}$) in the right-hand side of Eq.~(\ref{MK-alpha}) is independent of $\hat{m}$. This means the SOCC background effect, while effects due to the lens object come from the other terms.

\subsection{Self-contradiction in perturbative approximations in the literature}

First of all,
let us point out there is a self-contradiction in perturbative approximations
in \cite{Sultana, Cattani, Bhattacharya2010, Bhattacharya2011, Sultana2013},
which seems to have been unnoticed.

On deriving the deflection angle of light
in \cite{Sultana, Cattani, Bhattacharya2010, Bhattacharya2011, Sultana2013},
the MK metric is expanded around the Minkowski background.
This means
$|\gamma r| \ll 1$ in the metric expansion.

On the other hand,
the equation for a photon trajectory is expanded,
where the zeroth-order trajectory equation is assumed to
be in the form as
\cite{Sultana, Cattani, Bhattacharya2010, Bhattacharya2011, Sultana2013}
\begin{align}
\frac{d^2 u}{d\phi^2}
=
-u - \frac{\gamma}{2} ,
\label{zeroth}
\end{align}
and $u \equiv 1/r$.
This zeroth-order choice is a root of the problem.
We shall discuss this problem below.

If $|\gamma r| \ll 1$ is used for the zeroth-order equation,
the magnitude of the second term $- \gamma/2$ of the right-hand side
in Eq. (\ref{zeroth}) is much less than that of the first term $\sim 1/r$.
To be correct, therefore,
 the second term should be moved from the zeroth-order equation
to the next-order equation.

Namely, the use of the particular form of the zeroth-order equation
as Eq. (\ref{zeroth})
implicitly needs another assumption as $\gamma r = O(1)$.
Therefore,
there is a self-contradiction between
(1) $|\gamma r| \ll 1$ for the metric perturbation
and (2) $|\gamma r| = O(1)$ at the zeroth-order trajectory equation.
This inconsistency makes the previous results unconvincing.

One such example of the unconvincing results is that
the deflection angle of light in
\cite{Sultana, Cattani, Bhattacharya2010, Bhattacharya2011, Sultana2013}
includes an inverse-mass term.
See e.g. $-kb^3/(2\beta)$ of Eq.~(15) in \cite{Sultana},
$-k r_0^3/(2m)$ of Eq.~(17) in \cite{Cattani},
$- k R^3/(2m) $ of Eq.~(21) in \cite{Bhattacharya2010},
and
$-k R^3/(2m)$ in \cite{Sultana2013}
where $R = r_0$,
This inverse-mass term
diverges to infinity in the limit of the small lens mass ($m \to 0$).
If the inverse-mass term existed,
an infinitesimal mass could cause
a huge amount of the deflection of light.
However, this behavior makes no sense.
The unreasonable behavior is due to the inverse-mass term
as a consequence of the self-contradictory assumptions in the literature.

See Eq.~(\ref{MK-alpha}) for the deflection angle in the present paper.
There is no inverse-mass term in the deflection angle.
The present form of the deflection angle remains finite
in the zero mass limit ($m \to 0$).
Eq.~(\ref{MK-alpha}) is thus convincing.
This improvement is mainly because the SOCC method
incorporates the constant curvature $\hat{K}^{\mathrm{opt}}$
into background
and hence the iterative calculations are done self-consistently.
See e.g.~Eqs.~(\ref{OE-U}) and (\ref{B}).

We mention also minor issues.
First, we argue the allowed region  of the impact parameter.
$\alpha_{\mathrm{MK}}$ in Eq.~(\ref{MK-alpha}) is finite
since $\hat b$ is bounded by Eq.~(\ref{upperbound}),
though it apparently diverges in  $\hat b \to \infty$.
Namely, the limit of $\hat b \to \infty$ is not allowed in the SOCC formulation.
This point is contrast to the results in previous publications that are briefly summarized below
\cite{Edery, Pireaux, Sultana, Cattani, Bhattacharya2010, Bhattacharya2011, Sultana2013}.

Secondly, we shall argue a sign problem on $\gamma$.
At the direction $\phi = 0$,
which is indeed chosen for a coaligned setup,
the zeroth-order trajectory equation becomes
$u_0 = - \gamma/2$
\cite{Sultana, Cattani, Bhattacharya2010, Bhattacharya2011, Sultana2013}.
The left-hand side of this equation is positive
since $r_0 > 0$.
Hence $\gamma <0$.
However,
the literature \cite{Sultana, Cattani, Bhattacharya2010, Bhattacharya2011, Sultana2013} discusses the $\gamma > 0$ case by using the zeroth-order trajectory equation.
Therefore, there is another self-contradiction as opposite signs of $\gamma$.

Thirdly, an asymptotically nonflat issue is mentioned.
By using a conventional formulation appropriate for an asymptotically flat spacetime,
Reference \cite{Edery, Pireaux} find an extra deflection term as $-\gamma b$,
where $b \sim r_0$  is used in the weak deflection limit.
In their calculations, the receiver and source are located in the asymptotic infinity
($r \to \infty$).

In order to overcome the asymptotic nonflat problem,
Reference \cite{Sultana} employs the Rinder-Ishak method
to find from their linear-order calculations
extra terms as $- 2 \beta^2\gamma/b$ and $-k b^3/(2\beta)$
instead of $-\gamma b$. See Eq. (15) of Reference \cite{Sultana}.

After Reference \cite{Bhattacharya2010, Bhattacharya2011} argued
the deflection of light
to the second order in Rinder-Ishak method,
Reference \cite{Cattani} claimed
a new possible extra term in Eq. (17) as
$15 m^2 \gamma/b$,
whereas
Eq. (2.16) in Reference \cite{Sultana2013} insisted other types of extra terms
$15\pi m^2/(4b^2) + 37 m^3/(4b^3) + 6m^2 \gamma/b - k b^3/2m$.
The last term $- k b^3/2m$  in \cite{Sultana2013}
rapidly diverges as $b \to \infty$,
where we should stress that $b$ is unbounded in the Rindler-Ishak method.

Finally, we argue the Rindler-Ishak method,
in which the definition of angles
relies upon the spatial part of the metric.
However, the angle for light rays should be
defined by using the null geodesics.
In this regard,
the optical metric respects null geodesics.
The SOCC method thus uses the optical metric.

\subsection{Image position}
The relation between $U$ coordinate and $\rho$ coordinate
in hyperbolic background is found as
\begin{align}
  \label{hyperbolic-U}
  U &= \frac{\sqrt{-\hat{K}^{\mathrm{opt}}}}{\tanh\rho} \,\,.
\end{align}
Similarly, in spherical background, $U$ and $\rho$ are related as
\begin{align}
  \label{spherical-U}
  U &= \frac{\sqrt{\hat{K}^{\mathrm{opt}}}}{\tan\rho} \,\,.
\end{align}

In both Eqs.~(\ref{hyperbolic-U}) and (\ref{spherical-U}), the lens part of the deflection angle which is defined by Eq.~(\ref{alpha-lens}) is obtained in terms of $\theta$ as
\begin{align}
  \label{alpha-theta}
  \alpha(\hat m) = &\frac{4\hat{m}}{D_L \theta} - \frac{\hat{m}}{3 D_L}\left[1 + 3 \left(\frac{D_L}{D_{LS}}\right)^2\right]\theta \notag\\
  &+\frac{3}{2}\hat{m}\hat{\gamma}\left[1 + \frac{D_L}{D_{LS}} - \frac{1}{2}D_L^2 \hat{K}^{\mathrm{opt}} \left(1 + \frac{D_{LS}}{D_L}\right)\right]\theta \notag\\
  &- \hat{m} \hat{K}^{\mathrm{opt}} D_L \theta +\frac{3}{2}\hat m \hat \gamma \pi  +\mathcal{O}(\hat{m}^2) .
\end{align}

As already mentioned in Section $\mathrm{II}$, $\hat m / D_L \propto \varepsilon^2$ in our iterative method. Since $\varepsilon$ expansion of $\alpha$ is given by the expansion of $\theta$, the lowest order of $\alpha$ is $\mathcal{O}(\varepsilon)$.
By using Eq.~(\ref{expand-theta}) for Eq.~(\ref{alpha-theta}),
we obtain the expansion of
$\alpha$ up to the third-order of $\varepsilon$ as
\begin{align}
  \label{alpha-1st}
  \alpha_{(1)} &= \frac{4\hat{m}}{D_L \theta_{(1)}}\,\,,\\
  \label{alpha-2nd}
  \alpha_{(2)} &= -\frac{4\hat{m}}{D_L}\frac{\theta_{(2)}}{\theta_{(1)}^2} + \frac{3}{2}\hat{m}\hat{\gamma}\pi\,\,,
\end{align}
and
\begin{align}
  \label{alpha-3rd}
  \alpha_{(3)} = &-\frac{4\hat{m}}{D_L \theta_{(1)}} \left[\frac{\theta_{(3)}}{\theta_{(1)}} - \left(\frac{\theta_{(2)}}{\theta_{(1)}}\right)^2\right]\notag\\
  &- \frac{\hat{m}}{3 D_L}\left[1 + 3 \left(\frac{D_L}{D_{LS}}\right)^2\right]\theta_{(1)}\notag\\
  &-\hat{m}\hat K^{\mathrm{opt}} D_L \theta_{(1)}\notag\\
  &+\frac{3}{2}\hat{m}\hat{\gamma}\left(1 + \frac{D_L}{D_{LS}}\right)
  \left(1 - \frac{1}{2}D_L D_{LS} \hat K^{\mathrm{opt}}\right)\theta_{(1)}\,\,.
\end{align}

\subsection{Einstein ring}
In the $\varepsilon$ expansion, the Einstein ring radius can be formally expanded as
\begin{align}
  \theta_{E(n)} = \sum_{n=1}^{\infty} \varepsilon^{n}\theta_{E(n)}\,\,.
\end{align}
From Eqs.~(\ref{theta-1st})~-~(\ref{theta-3rd}) and (\ref{alpha-1st})~-~(\ref{alpha-3rd}), and a co-aligned setup $\beta = 0$, the Einstein radius is obtained up to the fourth-order of $\varepsilon$ as
\begin{align}
  \label{Einstein-1}
  \theta_{E(1)} &= \sqrt{\frac{4 \hat{m} D_{LS}}{D_L (D_L + D_{LS})}}\,\,,\\
  \label{Einstein-2}
  \theta_{E(2)} &= \frac{3}{16}D_L \theta_{E(1)}^2
 \hat{\gamma} \pi\,\,,
\end{align}
and
\begin{align}
  \label{Einstein-3}
  \theta_{E(3)} = &\frac{1}{2}\frac{\theta_{E(2)}^2}{\theta_{E(1)}} - \frac{1}{24}\left[1 + 3 \left(\frac{D_L}{D_{LS}}\right)^2 + 3 D_L^2 \hat K^{\mathrm{opt}}\right.\notag\\
  &- \frac{9}{2}D_L \hat{\gamma} \left(1 + \frac{D_L}{D_{LS}}\right) \left(1 - \frac{1}{2}D_L D_{LS} \hat K^{\mathrm{opt}}\right) \notag\\
  &\left.- \frac{2 D_L}{D_L + D_{LS}}\left(1 - \frac{D_L^2}{D_{LS}^2}\right)\right]\theta_{E(1)}^3\,\,.
\end{align}

\begin{figure}[ht]
  \includegraphics[width=8.6cm]{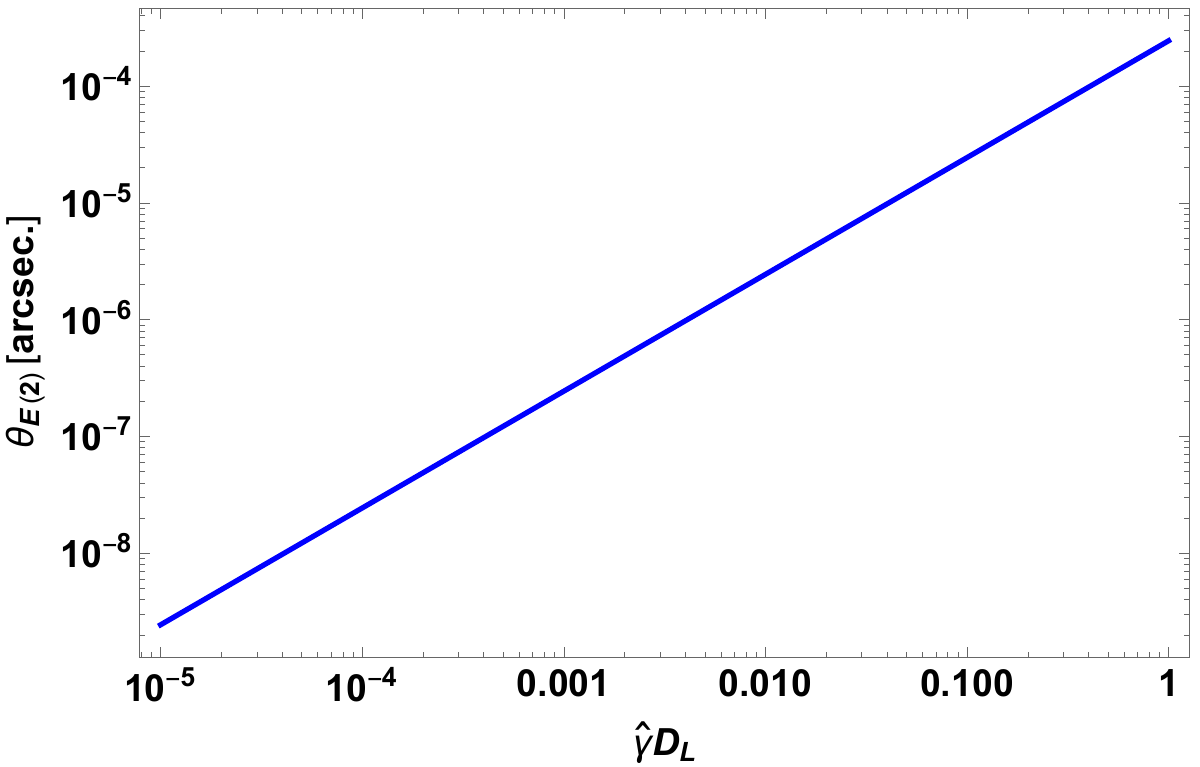}
  \caption{The plot of $\theta_{E(2)}$ in the galaxy scale. For its simplicity $D_L = D_{LS}$ is assumed. At the scale in the vicinity of $\hat \gamma D_L = \mathcal{O}(1)$, the correction by the $\hat \gamma$ parameter to the Einstein ring radius is of the order of 0.1 milli~arcseconds.}
  \label{Behav-Ering}
\end{figure}

Note that $\theta_{E(2)}$ does not appear in MK lensing by assuming the flat background~\cite{Takizawa2020b}. Eq.~(\ref{Einstein-2}) contains a coupling effect of the lens mass and the model parameter $\hat \gamma$. This is one important difference between the conventional method and the SOCC method. The constant-curvature term begins with $\theta_{E(3)}$, but it is unlikely to have relevance to the current observations.
If we assume a typical scale of the galaxy
($\hat m = 10^{12} M_{\odot}$) and $D_L = D_{LS} = 100\,\mathrm{Mpc}$, Eq.~(\ref{Einstein-3}) is estimated as $\theta_{(3)} \sim 10\, \mathrm{nano\,arcseconds}$,
where $\hat \gamma = 10^{-2}\,(\mathrm{Mpc})^{-1}$ and $\hat \gamma \gg \hat k$\, i.e.~$\hat K^{\mathrm{opt}} \sim -\hat \gamma^2$ are assumed. Henceforth we thus focus on the behavior of $\theta_{E(2)}$.
With the same parameter values, we can get $\theta_{E(2)} \sim 0.1\,\mathrm{milli\,arcseconds}$ which relate to the current VLBI observations if $|\hat \gamma| D_L \sim \mathcal{O}(1)$(Figure \ref{Behav-Ering}).

\subsection{Multiple images}

Next we consider $\beta \neq 0$ case. The image angle is obtained  at
the linear order of $\varepsilon$ from Eqs.~(\ref{theta-1st}) and (\ref{alpha-1st}) as
\begin{align}
  \label{image-1st}
  \theta_{(1)}^{\pm} = &\frac{1}{2} \frac{D_S}{D_L + D_{LS}}\notag\\ &\times \left[\beta_{(1)} \pm \sqrt{\beta_{(1)}^2 + 4\left(\frac{D_L + D_{LS}}{D_S}\right)^2 \theta_{E(1)}^2}\right]\,\,.
\end{align}
By substituting Eqs.~(\ref{alpha-2nd}) and (\ref{image-1st}) into Eq.~(\ref{theta-2nd}), the second-order image angle is obtained as
\begin{align}
  \label{image-2nd}
  \theta_{(2)}^{\pm} = 2 \left[1 + \frac{\theta_{E(1)}^2}{(\theta_{(1)}^{\pm})^2} \right]^{-1} \theta_{E(2)}\,\,,
\end{align}
where
\begin{align}
  \frac{3}{2}\hat m \hat \gamma \pi = 2 \frac{D_L + D_{LS}}{D_{LS}}\theta_{E(2)}\,\,,
\end{align}
is used for Eq.~(\ref{alpha-2nd}).

$\theta_{(2)}^{\pm}$ for primary and secondary images in Eq.~(\ref{image-2nd}) takes the same value. This means the same amount of translation of the both images.
However, the direction of the translation is opposite for $\theta_{(2)}^+$ and $\theta_{(2)}^-$ because the  primary and secondary images are produced in opposite sides with respect to the lens object. Namely, for the secondary image in Eq.~(\ref{image-2nd}), $\theta_{E(2)}$ in the right-hand side is replaced by $- \theta_{E(2)}$.

A separation angle between the primary and secondary images is $\Delta \theta \equiv |\theta^+ - \theta^-|$. In our iterative method, it can be written by series expansion in $\varepsilon$ as
\begin{align}
  \label{Separation-expand}
  \Delta \theta = &\sum_{n=1}^{\infty} \varepsilon^n \Delta\theta_{(n)}\notag\\
  =& \varepsilon |\theta_{(1)}^+ - \theta_{(1)}^-| + \varepsilon^2 |\theta_{(2)}^+ - \theta_{(2)}^-| + \mathcal{O}(\varepsilon^3)\,\,.
\end{align}
Let us examine a behavior of $\Delta\theta_{(2)}$, which can be calculated by using Eqs.~(\ref{image-1st}) and (\ref{image-2nd}) as
\begin{align}
  \label{Separation-2nd}
  \Delta\theta_{(2)} = 2 \theta_{E(2)}\,\,.
\end{align}
Therefore, $\Delta\theta_{(2)}$ is always constant.
For $\beta = 0$, we obtain $\Delta\theta_{(1)} + \Delta\theta_{(2)} = 2 (\theta_{E(1)} + \theta_{E(2)})$ which means a diameter of the Einstein ring. This result is consistent with calculations of the Einstein ring radius.

By assuming the galactic scale ($\hat m = 10^{12} M_{\odot}, D_L = D_{LS} = 100\,\mathrm{Mpc}$), the first-order Einstein ring radius is of the order of $\mathcal{O}(1)$ (arcseconds), where we assume $|\hat \gamma| D_L = \mathcal{O}(1)$ because the second-order Einstein ring radius can take a value which may relate to current observations. Then the magnitude of the second-order separation angle is estimated as $\Delta\theta_{(2)} \sim 0.1$ milli~arcseconds. This can be tested by current observations.

\section{Conclusion}

This paper extended the dS/AdS background method
based on the optical metric
to a SOCC background.
It was shown also that
the exact lens equation on the SOCC background
can be written in the same form as that for either Minkowski,
dS or AdS background
in terms of flat, spherical or hyperbolic trigonometry,
depending on the Gaussian curvature
of the equatorial plane in the SOCC background.

To exemplify the SOCC method,
we studied the gravitational lens in MK solution of Weyl gravity.
In the zero mass limit,
the deflection angle of light for the MK solution in the literature
diverges to infinity,
because there is a self-contradiction in perturbative approximations
of the MK metric and the orbit equation.
The SOCC method incorporates the long-distance curvature effect into the background.
Thereby the SOCC expression for the deflection angle of light in the MK solution
is finite also in the zero mass limit.

By solving the exact gravitational lens equation iteratively,
we found analytical solutions to the lens equation (i.e.~image positions) which differ from previous studies assuming the flat background.
For instance, coupling effects of $\gamma$ parameter on image positions
were found at the second order.
If $\hat\gamma$ parameter has a value such that $\hat\gamma D_L = \mathcal{O}(1)$,
the new term can
affect the Einstein ring radius at the order of 0.1 milli arcseconds in the galactic scale.

It is left for future to discuss the SOCC approach for a less symmetric case such as an axisymmetric spacetime.


\begin{acknowledgments}

We are grateful to Marcus Werner for carefully reading the earlier version of the manuscript
and providing useful comments on unclear points.
We thank Toshiaki Ono for the useful discussions
on the optical metric and the background subtraction.
We thank Hideyoshi Arakida and Yoo Chulmoon for the valuable discussions on the gravitational lens on non-flat background and the background dependence of the light deflection.
We wish to thank Mareki Honma for the conversations
on the EHT method and technology.
We thank Yuuiti Sendouda, Ryuichi Takahashi,
Masumi Kasai,  Ryuya Kudo, Yuya Nakamura,
Ryunosuke Kotaki, Masashi Shinoda, and  Hideaki Suzuki
for the useful conversations.
This work was supported
in part by Japan
Science and Technology Agency (JST) SPRING, Grant
Number, JPMJSP2152 (K.T.), and in part by Japan Society for the Promotion of Science (JSPS)
Grant-in-Aid for Scientific Research,
No. 20K03963 (H.A.),
and
in part by Ministry of Education, Culture, Sports, Science, and Technology,
Grant No. 17H06359 (H.A.)
\end{acknowledgments}


\begin{thebibliography}{99}
\bibitem{CCS}
   As is obvious from the spherical symmetry, the $\phi$ constant slice yields the same discussion. Moreover, since the metric depends only on the radial coordinate, the $r$ constant slice has a positive constant Gaussian curvature.
   Therefore, the three-dimensional Riemannian space described by the optical metric is a constant-curvature space. The sectional curvature is constant regardless of how to chose the hypersurface cut out.
\bibitem{footnote-1}
   If one introduces $\chi$ coordinate by $R \equiv \sin\chi$
   as usually done in cosmology instead of $\rho$,
   the proper length $L$ on the constant time slice is expressed as
   $dL^2 =  d\chi^2 + \sin\chi^2( d\Theta^2 + \sin^2 \Theta d\phi^2)$,
   with which cosmologists are familiar.
   The proper length $L$ describes a positive constant curvature space,
   namely spherical geometry.
   However, the proper length does not matter
   in the light propagation.
\end{thebibliography}
\end{document}